\documentclass[aps,superscriptaddress,preprintnumbers,showpacs,twocolumn]{revtex4}

\usepackage{amssymb}

\usepackage{indentfirst}

\usepackage{epsfig}

\usepackage{epsf}

\usepackage{graphicx}

\def\Vec#1{{\bf #1}}

\newcommand{\kp}{\mathbf{k}_\perp}

\newcommand{\p}{\perp}

\begin{document}

\newcommand*{\usm}{Departamento de F\'\i sica, Universidad T\'ecnica Federico
Santa Mar\'\i a, Casilla 110-V, Valpara\'\i so,
Chile}\affiliation{\usm}
\newcommand*{\pku}{School of Physics, Peking University, Beijing 100871, China}\affiliation{\pku}

\title{Transverse spin effects of sea quarks in unpolarized nucleons}

\author{Zhun Lu}\affiliation{\usm}
\author{Bo-Qiang Ma}\email[Corresponding author. Electronic address: ]{mabq@phy.pku.edu.cn}\affiliation{\pku}
\author{Ivan Schmidt}\email[Corresponding author. Electronic address: ]{ivan.schmidt@usm.cl}\affiliation{\usm}

\begin{abstract}
 We calculate the non-zero Boer-Mulders functions of
sea quarks inside the proton in a meson-baryon fluctuation model.
The results show that the transverse spin effects of sea quarks in
an unpolarized nucleon are sizable. Using the obtained antiquark
Boer-Mulders functions, we estimate the $\cos 2 \phi$ asymmetries in
the unpolarized $pp$ and $p D$ Drell-Yan processes at FNAL
E866/NuSea experiments. The prediction for the $\cos 2 \phi$
asymmetries in the unpolarized $pp$ Drell-Yan process at the BNL
Relativistic Heavy Ion Collider (RHIC) is also given.
\end{abstract}

\pacs{12.38.Bx, 13.85.-t, 13.85.Qk, 12.39.Ki }

\maketitle

\preprint{USM-TH-187}

\section{Introduction}

The transverse spin phenomena appearing in high energy scattering
processes~\cite{bdr} are among the most interesting issues of spin
physics. Several kinds of asymmetries associated with transverse
spin have been observed. One is the single spin asymmetry in the
semi-inclusive deeply inelastic scattering (SIDIS)
processes~\cite{smc,Airapetian:2004tw,compass,hermes05}, which
requires the target nucleon to be transversely polarized. Another is
the $\cos 2 \phi$ asymmetry in the unpolarized Drell-Yan
process~\cite{na10,conway}, where $\phi$ is the angle between the
lepton plane and the hadron plane. It has been demonstrated that
these asymmetries can be accounted for by specific leading-twist
$\Vec k_T$-dependent distribution functions, i.e., the Sivers
function~\cite{sivers,abm95} can explain~\cite{bhs02} the single
spin asymmetry in SIDIS processes, while the Boer-Mulders
function~\cite{bm} can account for~\cite{boer} the $\cos 2 \phi$
asymmetry in unpolarized Drell-Yan processes. Despite their naive
time reversal-odd property~\cite{collins93},
calculations~\cite{bhs02,gg02,bbh03,yuan,bsy04,lm04a,lm05} have
shown that these functions can be non-zero, due to the
gauge-links~\cite{collins02,belitsky,bmp03} appearing in their
operator definitions. The factorization theorem~\cite{Jmy04,cm04}
involving $\Vec k_T$-dependent distribution functions has been
worked out recently.

Sivers or Boer-Mulders functions correspond to a correlation between
the transverse spin of the nucleon or the quark and the transverse
momentum of the quark inside the nucleon, respectively. Therefore,
the investigation on them can provide information on the transverse
spin property of the nucleon at the quark level. The experimental
study of the Sivers function requires an incident nucleon
transversely polarized, with the advantage that this function
couples with usual unpolarized distribution or fragmentation
functions, and therefore it can be extracted more easily from
experimental data. In the case of the Boer-Mulders function, it
always couples with itself or another chiral-odd function in the
hard scattering process. Its advantage, however, is that the spin
structure of hadrons can be studied without invoking beam or target
polarization. In fact, the $\cos 2\phi$ asymmetries due to the
Boer-Mulders function of valence quarks have been studied
theoretically for the unpolarized $p\bar{p}$~\cite{bbh03,gg05,blm06}
and $\pi N$ ~\cite{lm05,lms06} Drell-Yan processes, which can be
measured in the future PAX~\cite{pax} or COMPASS experiments,
respectively. Measurements of the $\cos 2\phi$ asymmetries in the
unpolarized $p N$ Drell-Yan process can be performed by the FNAL
E866/NuSea collaboration and at the BNL Relativistic Heavy Ion
Collider (RHIC)~\cite{rhic}. This process is dominated by the
annihilation of valence and sea quarks from the two incident
nucleons. In this work we will investigate the role of the
Boer-Mulders functions of the nucleon sea, and reveal their impact
on the $\cos 2\phi$ asymmetries in the unpolarized $p N$ Drell-Yan
process. We calculate the Boer-Mulders functions of the intrinsic
$\bar{u}$ and $\bar{d}$ inside the proton using a meson-baryon
fluctuation model. Based on the resulting sea quark Boer-Mulders
functions, we estimate the $\cos 2 \phi$ asymmetries in unpolarized
$pp$ and $p D$ Drell-Yan processes at E866, and the $\cos 2 \phi$
asymmetries in unpolarized $p p$ Drell-Yan process at RHIC,
respectively. The magnitude of the asymmetry is of several percent,
and is sensitive to the choice of the Boer-Mulders functions of
valence quarks. Therefore the investigation of the asymmetry in the
unpolarized $p N$ processes can not only provide information on the
Boer-Mulders functions of intrinsic sea quarks, but also on those of
valence quarks.

\section{Boer-Mulders functions of sea quarks inside the proton}

The intrinsic sea quark content of the nucleon~\cite{meson} is
important for understanding its parton structure. Both experimental
measurements and model calculations have shown that there is a
sizable proportion of intrinsic sea quarks inside the nucleon,
including $u\bar{u}$, $d\bar{d}$ and $s\bar{s}$ sea quarks, which
can not be explained by the naive quark model. A qualitatively
successful model that accounts for a number of significant features
of the intrinsic sea quarks is the meson-baryon fluctuation
model~\cite{bm96}, which is
 based on the light-cone
Fock state expansion of the nucleon. The main assumption of this
model is that the intrinsic sea quarks are multi-connected to the
valence quarks and can exist over a relatively long lifetime within
the nucleon bound state, therefore the intrinsic $q\bar{q}$ pairs
can arrange themselves together with the valence quarks of the
target nucleon into the most energetically-favored meson-baryon
fluctuations. It is easy to understand that the most important
fluctuations are most likely to be those closest to the energy shell
and thus have minimal invariant mass. Such fluctuations are
necessary part of any quantum-mechanical description of the hadronic
bound state in QCD and have also been incorporated into the cloudy
bag model and Skyrme solution to chiral theories. Therefore, in the
meson-baryon fluctuation model, the nucleon can fluctuate into an
intermediate two-body system of a baryon and a meson which are in
turn composite systems of quarks and gluons. For example, the proton
can fluctuate into a $n\pi^+$ or $\Delta^{++}\pi^-$ configuration,
and the $\bar{d}$ or $\bar{u}$ inside $\pi^+$ or $\pi^-$ can be
viewed as $\bar{d}$ or $\bar{u}$ distributed in the proton.

Based on the
 meson-baryon fluctuation model one
can also calculate the momentum distribution of the intrinsic sea
quarks inside the proton, from the following convolution form
\begin{eqnarray}
q^{in}(x)&=&\int_x^1 \frac{dy}{y} f_{M \big{/}BM}(y) q_M\left
(\frac{x}{y} \right),\label{fsea}
\end{eqnarray}
where $f_{M \big{/}BM}(y)$ is the probability of finding the meson
$M$ in the $BM$ state with light-cone momentum fraction $y$, and
$q_M\left (x/y \right)$ is the probability of finding a quark $q$ in
the meson $M$ with light-cone momentum fraction $x/y$. The function
$f_{M \big{/}BM}(y)$ can be determined from the form
\begin{equation}
f_{M \big{/}BM}(y)=\int d^2\Vec k_T \left |\Psi_{BM}(y,\Vec k_T)
\right |^2,
\end{equation}
where $\Psi_{BM}(x,\Vec k_T)$ is the light-cone two-body wave
function of the baryon-meson system, for which we will choose a
Gaussian or a power-law behavior,
\begin{eqnarray}
\Psi_{B M}(x, \Vec k_T) & = & A_D \textrm{exp}(-M^2/8\alpha_D^2),\label{gaussian}\\
\Psi_{B M}(x, \Vec k_T) & = & A_D^\prime (1+M^2/
\alpha_D^2)^{-P},\label{powerlaw}
\end{eqnarray}
respectively. Here $M^2 =(M_M^2 + \Vec k_T^2)/x + (M_B^2 + \Vec
k_T^2)/(1-x)$ is the invariant mass squared of the baryon-meson
system, $\alpha_D$ sets the characteristic internal momentum scale,
and $A_D$ or $A_D^\prime$ is the normalization constant and can be
fixed by
\begin{equation}
\int dx  d^2 \Vec k_T |\Psi_{B M}(x, \Vec k_T)|^2=1.
\end{equation}
We point out that Eqs. (\ref{gaussian}) and (\ref{powerlaw}) are
boost-invariant light-cone wavefunctions which emphasize
multi-parton configurations of minimal invariant mass.

\begin{figure*}
\begin{center}
\begin{minipage}[b]{1.0\textwidth}
\scalebox{0.9}{\includegraphics*[91pt,3pt][322pt,190pt]{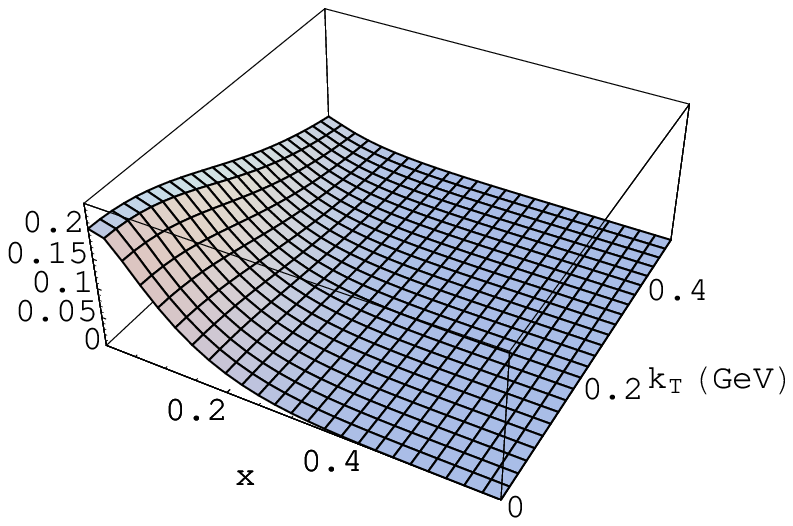}}
\scalebox{0.9}{\includegraphics*[91pt,3pt][322pt,190pt]{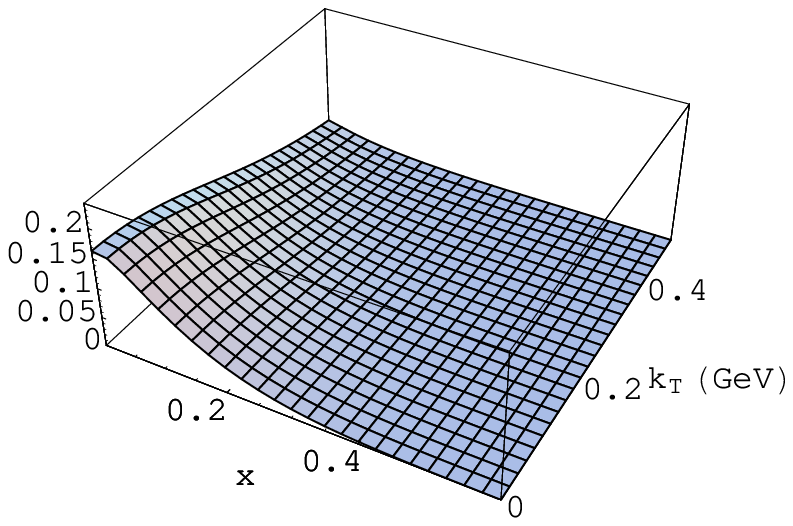}}
\scalebox{0.9}{\includegraphics*[91pt,3pt][322pt,190pt]{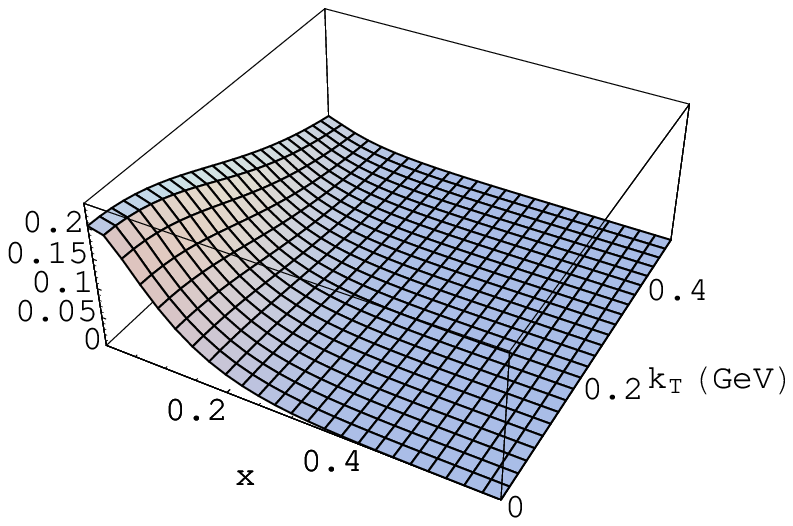}}
\scalebox{0.9}{\includegraphics*[91pt,3pt][322pt,190pt]{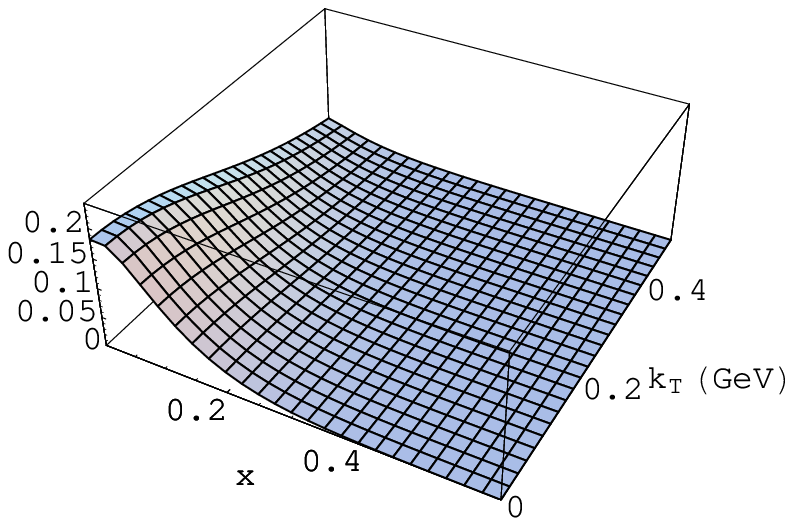}}
\end{minipage}

\caption{\small The Boer-Mulders functions of $\bar{u}$ quark
$h^{\perp,\bar{u}}_1(x,\Vec k_\perp^2)$ (left column) and $\bar{d}$
quark $h^{\perp,\bar{d}}_1(x,\Vec k_\perp^2)$ (right column) inside
the proton as two-dimensional densities. The upper panel and lower
panel correspond to the choice of {\rm gaussian} type and {\rm
power-low} type light-cone wave functions for the meson-baryon
system, respectively. }\label{udbar}
\end{center}
\end{figure*}

Eq.~(\ref{fsea}) can be extended to calculate the Boer-Mulders
functions of the intrinsic $\bar{u}$ and $\bar{d}$ antiquarks inside
the proton (denoted as $h^{\perp,\bar{u}}_1$ and
$h^{\perp,\bar{d}}_1$). In fact, according to the meson-baryon
fluctuation model, there are pion components in the intermediate
state of the proton. On the other hand, as shown in
Ref.~\cite{lm05}, non-zero Boer-Mulders functions of valence quarks
inside the pion (denoted by $h_{1\pi}^{\perp,q}$) can been
calculated using the quark-spectator-antiquark model. Therefore, a
convolution form similar to Eq.~(\ref{fsea}) can be applied to
calculate $h^{\perp,\bar{u}}_1$ and $h^{\perp,\bar{d}}_1$, as
follows:
\begin{eqnarray}
h^{\perp,\bar{u}}_1(x,\Vec k_\perp^2)&=&\int_x^1 \frac{dy}{y}
f_{\pi^- \big{/}\Delta^{++}\pi^-}(y) h^{\perp,\bar{u}}_{1\pi^-}\left
(\frac{x}{y},\Vec k_\perp^2 \right ),\label{hubar}\\
h^{\perp,\bar{d}}_1(x,\Vec k_\perp^2)&=&\int_x^1 \frac{dy}{y}
 f_{\pi^+\big{/}n\pi^+}(y)h^{\perp,\bar{d}}_{1\pi^+}\left
(\frac{x}{y},\Vec k_\perp^2 \right ).\label{hdbar}
\end{eqnarray}
The transverse momentum of the antiquark in the nucleon should be a
superposition of the transverse momentum of the antiquark in the
pion and of the transverse momentum of the pion in the nucleon. For
simplicity we neglect here the transverse momentum of the pion.
Assuming that the probabilities of the proton fluctuating to
$n\pi^+$ and $\Delta^{++}\pi^-$ (denoted as
$\mathcal{P}_{p\rightarrow n\pi^+}$ and $\mathcal{P}_{p\rightarrow
\Delta^{++}\pi^-}$, respectively) are both 12\%, and using the
functions $h^{\perp,q}_{1\pi}$ given in Ref.~\cite{lm05}, we obtain
the numerical results for $h^{\perp,\bar{u}}_1(x,k_T^2)$ and
$h^{\perp,\bar{d}}_1(x,k_T^2)$, shown in Fig.~\ref{udbar}. Two sets
of functions are given, corresponding to the choice of Gaussian
(upper panel) and power-law type (lower panel) wavefunctions of the
meson-baryon system, respectively. The parameters used in the
calculation are: $\alpha_D=0.33$ GeV, $P=3.5$ (as pointed out in
Ref.~\cite{bl80}, perturbative QCD predicts a nominal power-law fall
off at large $\Vec k_T$ corresponding to $P = 3.5$), following the
choice in Ref.~\cite{bm96}. The quark mass used in the calculation
is $m=0.3$ GeV. For the consistency here we use the same value for
the mass of the quark inside the proton and the pion, which is
different from that in Ref.~\cite{lm05} where we use 0.1 GeV for the
quark mass of the pion and 0.3 GeV for the quark mass of the proton.
This difference on the value of quark mass in the pion will bring a
quantitative difference on the numerical result, but the result of
the asymmetry and the main conclusion are qualitatively not changed.
It is interesting to point out that the Boer-Mulders function of
$\bar{u}$ or $\bar{d}$, calculated from the Gaussian type
wavefunction, qualitatively agrees with that from the power-law type
wave function. We should notice that this result is based on the
assumption that the numbers of the sea quark pairs $u\bar{u}$ and
$d\bar{d}$ inside the nucleon are the same, since we assume
$\mathcal{P}_{p\rightarrow n\pi^+} = \mathcal{P}_{p\rightarrow
\Delta^{++}\pi^-}$ here. As we know that there should be an excess
of $d\bar{d}$ pair over $u\bar{u}$ pair in the proton, which is a
consequence of Gottfried sum rule~\cite{gsm}, in next section we
will apply another set of $\mathcal{P}_{p\rightarrow n\pi^+}$ and $
\mathcal{P}_{p\rightarrow \Delta^{++}\pi^-}$, which is constrained
by the existing parametrization that can produce the flavor
asymmetry of the sea quarks, and investigate the consequent
asymmetries in the unpolarized $p p$ and $p D$ Drell-Yan processes.
Boer-Mulders functions for sea quarks may be obtained from other
models that also allow sea quarks in the nucleon, such as the chiral
quark and meson cloud models. In these models there are also $\pi$
meson components in the Fock state expansion of the nucleon.

\section{The $\cos 2\phi$ asymmetries in the unpolarized $pp$ and $pD$ Drell-Yan processes}

In this section, we will calculate the $\cos 2\phi$ asymmetries in
the unpolarized $pp$ and $pD$ Drell-Yan processes, based on the
Boer-Mulders functions of sea quarks obtained above. The general
form of the angular differential cross section for the unpolarized
Drell-Yan process is
\begin{eqnarray}
\frac{1}{\sigma}\frac{d\sigma}{d\Omega}&=&\frac{3}{4\pi}\frac{1}{\lambda+3}
\left
(1+\lambda\textmd{cos}^2\theta+\mu\textmd{sin}2\theta\textmd{cos}\phi
\right.
\nonumber\\
& & \left .
+\frac{\nu}{2}\textmd{sin}^2\theta\textmd{cos}2\phi\right
).\label{cos2phi}
\end{eqnarray}

The $\phi$ independent differential cross-section for the
unpolarized $pp$ Drell-Yan process is
\begin{eqnarray}
\frac{d\sigma(pp\rightarrow l\bar{l}X)}{d\Omega
dx_1dx_2d^2\mathbf{q}_\perp}&=&\frac{\alpha_{em}^2}{12Q^2}(1+\cos^2\theta)
\sum_{q=u,d}e_q^2 \mathcal{F}[f_1^q(x_1,\mathbf{p}_\perp^2)
\nonumber\\
&&\times f_{1}^{\bar{q}}(x_2,\mathbf{k}_\perp^2)]+(q \leftrightarrow
\bar{q}),\label{phidep}
\end{eqnarray}
where we have used the notation~\cite{cs77}:
\begin{equation}
\mathcal{F}[\cdots]=\int d^2\mathbf{p}_\perp
d^2\kp\delta^2(\mathbf{p}_\perp+\kp-\mathbf{q}_\perp)\times\{\cdots\},
\end{equation}
and the transverse momentum dependence of the cross-section is
implicit.

The corresponding cross section for the unpolarized $pD$ Drell-Yan
process is then
\begin{eqnarray}
\frac{d\sigma(pD\rightarrow l\bar{l}X)}{d\Omega
dx_1dx_2d^2\mathbf{q}_\perp}&=&\frac{\alpha_{em}^2}{12Q^2}(1+\cos^2\theta)
\mathcal{F}[(e_u^2
f_1^u(x_1,\mathbf{p}_\perp^2)\nonumber\\
&&+e_d^2f_1^d(x_1,\mathbf{p}_\perp^2))\times(f_{1}^{\bar{u}}(x_2,\mathbf{k}_\perp^2)
\nonumber\\
&&+f_{1}^{\bar{d}}(x_2,\mathbf{k}_\perp^2))] +(q \leftrightarrow
\bar{q}),\label{2phidep}
\end{eqnarray}
Where we have used the isospin relation $f_1^{u/D} \approx
f_1^{u/p}+f_1^{u/n}=f_1^u+f_1^d$.

Eqs. (\ref{phidep}) and (\ref{2phidep}) are lowest order parton
model expressions, where it is extended in order to include
transverse parton momenta.

The $\cos2\phi$ dependent cross-sections for unpolarized $p p$ and
$p D$ Drell-Yan processes, in terms of Boer-Mulders functions, are
\begin{eqnarray}
&&\left.\frac{d\sigma(pp\rightarrow l\bar{l}X)}{d\Omega
dx_1dx_2d^2\mathbf{q}_\perp}\right
|_{\cos2\phi}\nonumber\\
\hspace{-2cm}&=&\frac{\alpha_{em}^2}{12Q^2}
\sin^2\theta\cos2\phi\sum_{q=u,d}e_q^2
\mathcal{F}[\chi(\mathbf{p}_\perp,\mathbf{k}_\perp) h_{1}^{\perp,q}(x_1,\mathbf{p}_\perp^2)\nonumber\\
&&\times h_{1}^{\perp,\bar{q}}(x_2,\mathbf{k}_\perp^2)]+(q \leftrightarrow \bar{q});\\
&&\left.\frac{d\sigma(pD\rightarrow l\bar{l}X)}{d\Omega
dx_1dx_2d^2\mathbf{q}_\perp}\right
|_{\cos2\phi}\nonumber\\
&=&\frac{\alpha_{em}^2}{12Q^2} \sin^2\theta\cos2\phi \,
\mathcal{F}[\chi(\mathbf{p}_\perp,\mathbf{k}_\perp)
(e_u^2 h_1^{\perp,u}(x_1,\mathbf{p}_\perp^2)\nonumber\\
&&+e_d^2 h_1^{\perp, d}(x_1,\mathbf{p}_\perp^2))
(h_{1}^{\perp,\bar{u}}(x_2,\mathbf{k}_\perp^2)
+h_{1}^{\perp,\bar{d}}(x_2,\mathbf{k}_\perp^2)]\nonumber\\
&&+(q \leftrightarrow \bar{q}),
\end{eqnarray}
respectively, where
\begin{equation}
\chi(\mathbf{p}_\perp,\mathbf{k}_\perp)=(2\hat{\mathbf{h}}\cdot
\mathbf{p}_\perp\hat{\mathbf{h}}\cdot \mathbf{k}_\perp
-\mathbf{p}_\perp\cdot \kp)/M_p^2,
\end{equation}
here $\hat{\mathbf{h}}=\Vec q_T/Q_T$ and $M_p$ is the mass of the
proton.

Therefore we can express the $\cos2\phi$ asymmetry coefficient
defined in Eq.~(\ref{cos2phi}) as ($\lambda=1$, $\mu=0$)
\begin{eqnarray}
\nu_p&=&\frac{2\sum_{q=u,d}e_q^2 \mathcal{F}[\chi h_{1}^{\perp,q}
h_1^{\perp,\bar{q}} ]+ (q \leftrightarrow
\bar{q})}{\sum_{q=u,d}e_q^2 \mathcal{F}[f_1^q f_{1}^{\bar{q}}]+
(q \leftrightarrow \bar{q})},\label{nup}\\
\nu_D&=&\frac{2\mathcal{F}[\chi (e_u^2 h_1^{\perp,u} +e_d^2
h_1^{\perp, d}) (h_{1}^{\perp,\bar{u}} +h_{1}^{\perp,\bar{d}})]+(q
\leftrightarrow \bar{q})}{\mathcal{F}[ (e_u^2 f_1^{\perp,u} +e_d^2
f_1^{\perp, d}) (f_{1}^{\perp,\bar{u}} +f_{1}^{\perp,\bar{d}})]+(q
\leftrightarrow \bar{q})},\nonumber\\ \label{nud}
\end{eqnarray}
where we have omitted the arguments of the distribution functions.

Eqs.~(\ref{nup}) and (\ref{nud}) give the explanation for the $\cos
2 \phi$ asymmetry observed in the unpolarized Drell-Yan process from
the view of Boer-Mulders function. We note that there are other
theoretical approaches that have been proposed to interpret this
asymmetry, such as high-twist~\cite{bbkd94,ehvv94} and QCD vacuum
effects~\cite{bnm93}. In Ref.~\cite{bbnu} detailed comparison
between the QCD Vacuum effect and Boer-Mulers effect has been done.

The E866 Collaboration at FNAL employs a 800 GeV/c proton beam
colliding on protons or deuterons fixed targets to measure the
unpolarized Drell-Yan processes $pp\rightarrow \mu^+\mu^- X$ and
$pD\rightarrow \mu^+\mu^- X$~\cite{e866_dy}. The main goal of E866
is to study the asymmetry of the nucleon sea
distribution~\cite{e866_sea1,e866_sea2}. These experiments can also
be applied to study the $\cos 2 \phi$ asymmetry of the lepton pair,
especially the role of the sea quarks contribution to the asymmetry.
In the following we will estimate the asymmetry $\nu_p$ and $\nu_D$
for the E866 experiment. In order to calculate the asymmetry we also
need the specific form of Boer-Mulders functions of valence quarks
inside the nucleon. Since there are no experimental measurements on
those functions yet, we use some model results for them, such as
those given in Ref.~\cite{bsy04}, within a quark and spectator
diquark model of the nucleon. We will take two different extreme
options as follows.
\begin{itemize}
    \item  Option I: Consider the limit with only the spectator scalar diquark
    included, which means that $h_1^{\p,u}=h_1^{\p,S}$ and
    $h_1^{\p,d}=0$. Here we use $h_1^{\p,S}$ and $h_1^{\p,V}$ to represent the contributions to the
    Boer-Mulders functions with only spectator
    scalar diquark and vector diquark included, respectively. This option was
    also applied in Ref.~\cite{lm05}
    \item  Option II: Consider the case with both the spectator scalar diquark and vector
    diquark included. In this option $h_1^{\p,u}=\frac{3}{2}h_1^{\p,S}+\frac{1}{2}h_1^{\p,V}$ and
    $h_1^{\p,d}=h_1^{\p,V}$, and they have opposite sign.
\end{itemize}
The scalar and vector components of the proton are~\cite{bsy04}
\begin{eqnarray}
 h_1^{\p,S}(x,\Vec k_T^2) & =& \frac{4}{3} \,  \alpha_s \, N_s\, (1 -
x)^{3} \, \frac{M \, (x M + m)}{[L_s^2 \, (L_s^2 + \Vec k_T^2)^3]},
\label{h1s} \\
 h_1^{\p,V}(x, \Vec k_T^2)& =& - \frac{4}{3} \,  \alpha_s \, N_v\, (1 -
x)^{3} \, \frac{M \, (2x M + m)}{[L_v^2 \, (L_v^2 + \Vec k_T^2)^3]},
\label{h1v}
\end{eqnarray}
respectively, where $N_{s/v}$ is a normalization constant, and
\begin{equation}
L^2_{s/v}=(1 - x) \, \Lambda^2 + x \, M_{s/v}^2  - x \, (1 - x) \,
M^2\,.
\end{equation}
Here $\Lambda$ is a cutoff appearing in the nucleon-quark-diquark
vertex and $M_d$ is the mass of the scalar diquark.

\begin{figure}
\begin{center}
\scalebox{0.8}{\includegraphics*{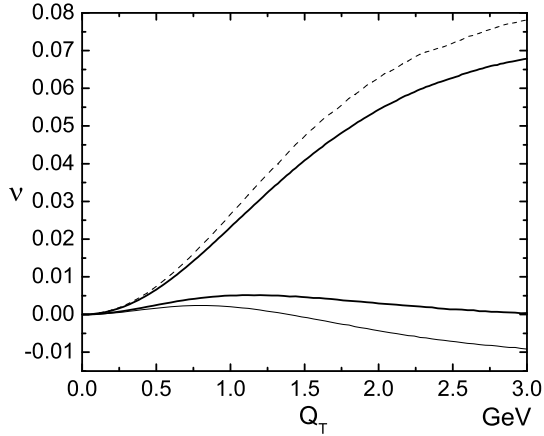}}
 \caption{\small
The $\cos 2 \phi$ asymmetries $\nu_p$ (thick lines) and $\nu_D$
(thin lines) for the E866 experiment, as functions of $Q_T$, with
kinematical cuts $4.5 \,\textrm{GeV} < Q < 9 \,\textrm{GeV}$ and $
Q>11 \,\textrm{GeV}$, $0.1< x_1 <1$, $0.015<x_2<0.4$. The dashed and
solid lines show the asymmetries calculated from options I and II,
respectively.}\label{e866}
\end{center}
\end{figure}

With the following kinematical cuts
\begin{eqnarray}
&&4.5 \,\textrm{GeV} < Q < 9 \,\textrm{GeV} ~~~~\textrm{and}~~~ Q>11 \,\textrm{GeV},\nonumber\\
&&0.1< x_1 <1,~~~~0.015<x_2<0.4,\nonumber
\end{eqnarray}
we calculate the $Q_T$-dependent $\cos 2 \phi$ asymmetries in the
unpolarized $pp$ and $pD$ Drell-Yan processes at E866, as shown in
Fig.~\ref{e866}. In the calculation we adopt the antiquark
Boer-Mulders functions $h_1^{\p,\bar{q}}$ calculated from the
Gaussian type wavefunction of the baryon-meson system.
One observation from Fig.~\ref{e866} is that the asymmetry
calculated according to option I is of several percent, while that
according to option II is around one percent, indicating that the
size of the asymmetry is sensitive to the choice of the Boer-Mulders
functions of valence quarks. Nevertheless, our calculation predicts
a quite smaller $\cos 2 \phi$ asymmetry in the unpolarized $pN$
Drell-Yan process compared to that in the unpolarized $\pi N$
Drell-Yan case~\cite{na10,lm05}. Again, comparing the asymmetries
calculated from the two different options, we find that $\nu_p$ and
$\nu_D$ have similar sizes in option I, while those of $\nu_p$ and
$\nu_D$ in option II can be very different, and even have opposite
signs. Therefore, the investigation of the $\cos 2 \phi$ asymmetry
in the unpolarized $p N$ Drell-Yan process, can not only give
information on the Boer-Mulders functions of sea quarks, but also
about those of valence quarks.

\begin{figure}
\begin{center}
\scalebox{0.9}{\includegraphics[10pt,15pt][275pt,230pt]{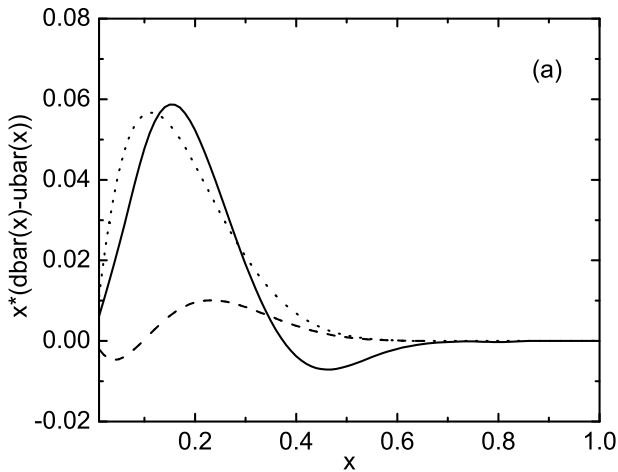}}
\caption{\small  The flavor asymmetry of the sea quark distributions
$x(\bar{d}(x)-\bar{u}(x))$. The solid curve shows the result from
CTEQ6L parametrization at $Q=0.6$ GeV. The dashed and dotted line
correspond to the results from Set I and II,
respectively.}\label{unp}
\end{center}
\end{figure}

\begin{figure}
\begin{center}
\scalebox{0.8}{\includegraphics{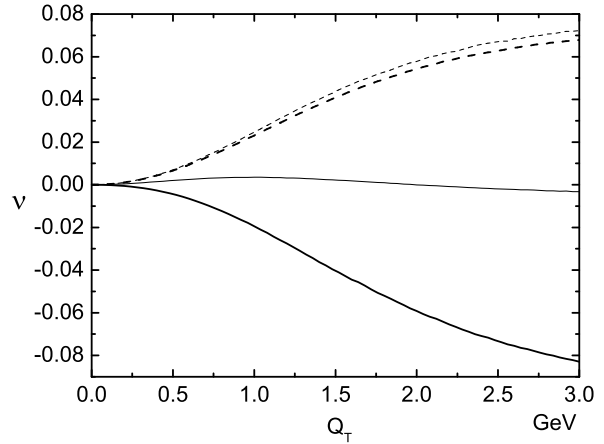}} \caption{\small Same
as Fig. ~\ref{e866}, but using Set II for the sea quark
distributions which produce the flavor asymmetry of sea quark
indicated by the Gottfried sum rule.}\label{e866norm}
\end{center}
\end{figure}

The main success of the baryon-meson fluctuation model is that it
can qualitatively describe a variety of asymmetries of the nucleon
sea distributions, such as the flavor asymmetry
$\bar{d}(x)-\bar{u}(x)$, the strange sea asymmetry
$s(x)-\bar{s}(x)$, and the polarized sea asymmetry $\Delta
s(x)-\Delta \bar{s}(x)$. In the previous section we have assumed
that the probabilities of the proton fluctuating to $n\pi^+$ and
$\Delta^{++}\pi^-$ are 12\% (here we denote it as Set I for the sea
quark distributions). From this set we calculate the flavor
asymmetry $x(\bar{d}-\bar{u})$, as shown by the dashed line in
Fig.~\ref{unp}, and compare it with that from the CTEQ6L
parametrization at Q=0.6 GeV (shown by the solid line in the same
figure). None-zero asymmetry of the sea quark densities is found in
this case, but is quantitatively much lower than the known
parametrization. In a further consideration, we will take into
account the flavor asymmetry of the sea quark and adopt
$\mathcal{P}_{p\rightarrow n\pi^+}=15\%$ and
$\mathcal{P}_{p\rightarrow \Delta^{++}\pi^-}=1\%$ (denoted as Set
II), motivated by the expectation that there should be an excess of
$u\bar{u}$ pair over $d\bar{d}$ pair, which is a consequence of
Gottfried sum rule~\cite{gsm}. This set has also been adopted in
Ref.~\cite{msy}. We show the asymmetry $x(\bar{d}-\bar{u})$ from Set
II by the dotted line in Fig.~\ref{unp}, and find that it agrees
with the flavor asymmetry from CTEQ6L parametrization fairly well.
From the sea quark distributions of Set II, we give the $\cos 2
\phi$ asymmetries in the unpolarized $pp$ (thick line) and $p D$
(thin line) Drell-Yan process at E866, as shown in
Fig.~\ref{e866norm}. Two options of the valence quark distribution
are applied in this calculation. Again, the asymmetries $\nu_p$ and
$\nu_D$ have similar sizes in option I, and those in option II are
very different. In Fig.~\ref{e866norm}, a larger negative $\nu_p$
from option II is predicted compared to that given in
Fig~\ref{e866}.

As we don't know the exact form of Boer-Mulders functions of valence
quarks, we admit that our prediction, in a certain aspect, relies on
functions given in \cite{bsy04}, where the Boer-Mulders functions
for the $u$ and $d$ quarks are of the opposite sign. Actually
different models or theoretical considerations predict different
flavor dependence of valence Boer-Mulders functions. A calculation
based on the MIT bag model gave~\cite{yuan}
$h_1^{\perp,d}=\frac{1}{2}h_1^{\perp,u}$, while the argument based
on large-$N_c$ showed that
$h_1^{\perp,d}=h_1^{\perp,u}$~\cite{pobylista} modulo $1/N_c$
corrections. Recent lattice work~\cite{lattice} and
works~\cite{gpds} on impact parameter representation of generalized
parton distributions (GPDs) state that $h_1^{\perp,u}$ and
$h_1^{\perp,d}$ are of the same sign. In those cases the asymmetries
may be different from the prediction given in this work. In our
recent paper~\cite{lms06}, it has been suggested to study the flavor
dependence of valence Boer-Mulders functions, through $\pi p$ and
$\pi D$ Drell-Yan processes that can be performed in the coming
years by COMPASS Collaboration. In this paper, We have taken two
options for the valence Boer-Mulders functions to study how the
asymmetry depends on them. Nevertheless, the measurement of the
$\cos 2 \phi$ asymmetries in the unpolarized $p p$ and $p D$
processes may provide useful constraint on the Boer-Mulders function
of valence quarks.

\begin{figure}
\begin{center}
\scalebox{0.8}{\includegraphics{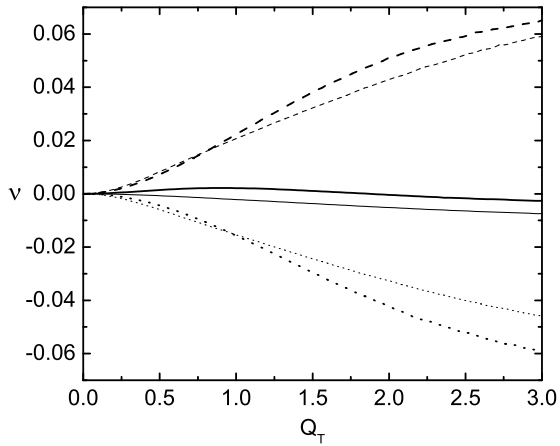}} \caption{\small The
$Q_T$-dependent $\cos 2 \phi$ asymmetries $\nu_p$  at RHIC, with the
kinematical conditions $\sqrt{s}=200$ GeV, $-1< y <2$. The thin  and
thick lines show the asymmetry at $Q=4$ and 20 GeV, respectively.
The dashed lines show the asymmetry from option I and Set I or II
(in option one the asymmetries in Set I and II are the same), and
solid lines show the asymmetries calculated from options II and Set
I, dotted lines show the asymmetries calculated from options II and
Set II, respectively.}\label{rhic}
\end{center}
\end{figure}

Another promising test ground of $\cos 2 \phi$ asymmetries in the
unpolarized $pN$ Drell-Yan process is BNL RHIC~\cite{rhic}, where
the Drell-Yan process $pp\rightarrow l^+l^- X$ can be measured.
There are already numerous
studies\cite{boer,soffer02,anselmino03,collins06} on the single
spin~ and double spin asymmetries in Drell-Yan processes at RHIC,
which require at least one incident proton to be polarized. The
unpolarized experiment is also feasible at RHIC. Taking the same
sets of functions (Set I and II for sea quark distributions and
Option I and II for valence quark distributions) used to calculate
the asymmetries at E866, we estimate the $\cos 2 \phi$ asymmetry
$\nu_p$ at RHIC with the kinematical constraints $\sqrt{s}=200$ GeV,
$-1 < y <2$, here $y$ is the rapidity defined as
$y=\frac{1}{2}\textrm{ln}\frac{x_1}{x_2}$. The $Q_T$-dependent
asymmetries for $Q=4$ (thin line) and $Q=20$ GeV (thick line) are
shown in Fig.~\ref{rhic}. Here we do not average the asymmetry over
a range of $Q$, as we have done for the asymmetry at E866, since we
would like to choose two different fixed values of $Q$ to study the
$Q$-dependence of the asymmetry at RHIC. Since the evolution
equations of unintegrated parton distributions, especially that of
the Boer-Mulders functions, are unknown, we assume that they scale
with $Q$. Therefore the different asymmetries at different $Q$
values, as shown in Fig.~\ref{rhic}, is not a consequence of
evolution, but just a kinematical effect.

\section{Conclusion}
In summary, we have applied a meson-baryon fluctuation model to
calculate the Boer-Mulders functions of $\bar{u}$ and $\bar{d}$
inside the proton, and provided a first estimate of the transverse
spin effect of sea quarks in the unpolarized nucleon. In the
calculation we adopted both the Gaussian and the power-law type
wavefunctions for the meson-baryon system. We found the sizes of
$h_1^{\p,\bar{u}}$ or $h_1^{\p,\bar{d}}$ from these two types of
wavefunctions to be qualitatively similar. From the obtained
antiquark Boer-Mulders functions, we analyzed the $\cos 2 \phi$
asymmetries in the unpolarized $p p$ and $p D$ Drell-Yan processes
at E866. Two sets of the sea quark distributions are applied in the
calculation, in which Set I is flavor symmetric and Set II reflects
the sea quark flavor asymmetry indicated by Gottfried sum rule. The
size of the asymmetries is at most several percent, in the case that
the Boer-Mulders functions of valence quarks are chosen with only
spectator scalar-diquark contributed. The size of the asymmetries,
as well as the relative size between the asymmetries in the
unpolarized $p p$ and $p D$ Drell-Yan processes, relies
significantly on the choice of Boer-Mulders functions of valence
quarks. We also estimated the $\cos 2\phi$ asymmetry in the
unpolarized $pp$ Drell-Yan process at RHIC.  This investigation
suggests that unpolarized $pN$ Drell-Yan processes on hadron
colliders are very helpful for a better understanding about the role
of transverse spin of both valence and sea quarks in the unpolarized
nucleon.

{\bf Acknowledgements.} We acknowledge helpful discussions with
J.C.~Peng and L.~Zhu. This work is partially supported by National
Natural Science Foundation of China (Nos.~10421003, 10575003,
10505011, 10528510), by the Key Grant Project of Chinese Ministry of
Education (No.~305001), by the Research Fund for the Doctoral
Program of Higher Education (China), by Fondecyt (Chile) under
Project No.~3050047 and No.~1030355.


\begin{thebibliography}{99}

\bibitem{bdr}
For a review on tranverse polarization phenomena, see V.~Barone,
A.~Drago, P.G.~Ratcliffe, Phys. Rep. {\bf 359}, 1 (2002).

\bibitem{smc} A. Bravar et al., SMC Collaboration, Nucl. Phys. A {\bf 666}, 314 (2000) .

\bibitem{Airapetian:2004tw}

A.~Airapetian et al., HERMES Collaboration, Phys. Rev. Lett. {\bf
94}, 012002 (2005) .

\bibitem{compass}

V.Yu.~Alexakhin et al., COMPASS Collaboration, Phys. Rev. Lett. {\bf
94}, 202002 (2005).

\bibitem{hermes05} M. Diefenthaler, HERMES Collaboration,
 in Proceedings of DIS 2005, Madison, Wisconsin (USA), hep-ex/0507013.

\bibitem{na10} S.~Falciano et al. NA10 Collaboration,
Z. Phys. C {\bf 31}, 513 (1986); M.~Guanziroli et al. NA10
Collaboration, Z. Phys. C {\bf 37}, 545 (1988).

\bibitem{conway} J.S.~Conway et al.
Phys. Rev. D {\bf 39}, 92 (1989).

\bibitem{sivers} D. Sivers, Phys. Rev. D {\bf 41}, 83 (1990);
D. Sivers, Phys. Rev. D {\bf 43}, 261 (1991).

\bibitem{abm95} M. Anselmino, M. Boglione, F. Murgia,
Phys. Lett.  B {\bf 362}, 164 (1995).

\bibitem{bhs02} S.J.~Brodsky, D.S.~Hwang,
 I.~Schmidt, Phys. Lett. B {\bf 530}, 99 (2002).

\bibitem{bm}

D.~Boer, P.J.~Mulders, Phys. Rev. D {\bf 57}, 5780 (1998).

\bibitem{boer}

D.~Boer, Phys. Rev. D {\bf 60}, 014012 (1999).

\bibitem{collins93} J.C.~Collins, Nucl. Phys. B {\bf 396}, 161 (1993).

\bibitem{gg02}  G.R. Goldstein, L. Gamberg, Talk given at 31st International
Conference on High Energy Physics (ICHEP 2002), Amsterdam, The
Netherlands, 24-31 July 2002, hep-ph/0209085.

\bibitem{bbh03} D.~Boer, S.J.~Brodsky, D.S.~Hwang, Phys.
Rev. D {\bf 67}, 054003 (2003).

\bibitem{yuan}
F. Yuan, Phys. Lett. B {\bf 575}, 45 (2003).

\bibitem{bsy04} A.~Bacchetta, A.~Sch\"{a}fer,
 J.-J.~Yang, Phys. Lett. B {\bf 578}, 109 (2004).

\bibitem{lm04a} Z.~Lu and B.-Q.~Ma, Nucl. Phys. A {\bf 741}, 200
(2004); Z.~Lu, B.-Q.~Ma, Phys. Rev. D {\bf 70}, 094044 (2004).

\bibitem{lm05}
Z.~Lu, B.-Q.~Ma, Phys. Lett. B {\bf 615}, 200 (2005).

\bibitem{collins02} J.C. Collins,  Phys. Lett. B {\bf 536}, 43 (2002).

\bibitem{belitsky}
X.~Ji and F.~Yuan, Phys. Lett. B {\bf 543}, 66 (2002);
A.V.~Belitsky, X.~Ji, F.~Yuan, Nucl. Phys. B {\bf 656}, 165 (2003).



\bibitem{bmp03} D. Boer, P.J. Mulders, and F. Pijlman, Nucl. Phys. B {\bf 667}, 201 (2003).

\bibitem{Jmy04}

X.~Ji, J.P.~Ma and F.~Yuan, Phys. Lett.  B {\bf 597}, 299 (2004).

\bibitem{cm04} J.C. Collins, A. Metz, Phys. Rev. Lett. {\bf 93},
252001 (2004).

\bibitem{gg05} G. R. Goldstein, L. Gamberg, hep-ph/0506127.

\bibitem{blm06} V. Barone, Z.~Lu and B.-Q.~Ma, to appear in European
physical journal.

\bibitem{lms06} Z. Lu, B.-Q. Ma and I. Schmidt, Phys. Lett. B {\bf
639} 494 (2006).


\bibitem{pax}
PAX Collaboration, V.~Barone, et al., hep-ex/0505054.

\bibitem{rhic} G. Bunce, N. Saito, J. Soffer and W. Vogelsang,
Ann. Rev. Nucl. Part. Sci. {\bf 50}, 525 (2000).

\bibitem{meson} For reviews, see, e.g., S. Kumano, Phys. Rep. 303
(1998) 183;\\
G.T. Garvey, J.C. Peng, Prog. Part. Nucl. Phys. 47 (2001) 203.

\bibitem{bm96} S.J. Brodsky and B.-Q. Ma, Phys. Lett. B {\bf 381}, 317 (1996).


\bibitem{bl80} G.P. Lepage and S.J. Brodsky, Phys. Rev. D {\bf 22},
2157 (1980).

\bibitem{gsm} K. Gottfried, Phys. Rev. Lett. {\bf 18}, 1174 (1967).

\bibitem{cs77} J.C. Collins and D.E. Soper, Phys. Rev. D {\bf 16},
2219 (1977).

\bibitem{bbkd94} A. Brandenburg, S.J. Brodsky, V.V. Khoze, and D.
M\"{u}ller, Phys. Rev. Lett. {\bf 73}, 939 (1994).

\bibitem{ehvv94} K.J. Eskola, P. Hoyer, M. V\"{a}nttinen, and
R. Vogt, Phys. Lett. {\bf B333}, 526 (1994).

\bibitem{bnm93} A.
Brandenburg, O. Nachtmann, and E. Mirkes, Z. Phys. C {\bf 60}, 697
(1993).

\bibitem{bbnu} D. Boer, A. Brandenburg, O. Nachtmann and A. Utermann, Eur.
Phys. J. C {\bf 40} 55 (2005) .

\bibitem{e866_dy} J.C. Webb {\it et al.}, (E866 Collaboration), hep-ex/0302019.

\bibitem{e866_sea1} E.A. Hawker {\it et al.}, (FNAL E866/NuSea Collaboration), Phys. Rev.
Lett. {\bf 80}, 3715 (1998).

\bibitem{e866_sea2} R.S. Towell {\it et al.}, (FNAL
E866/NuSea Collaboration), Phys. Rev. D {\bf 64}, 052002 (2001).

\bibitem{msy} B.-Q. Ma, I. Schmidt and J.-J. Yang, Eur. Phys. J. A
{\bf 12}, 353 (2001).



\bibitem{pobylista} P.V. Pobylitsa, hep-ph/0301236.

\bibitem{lattice} M. Gockeler {\it et al.}, Nucl. Phys. Proc. Suppl. {\bf 153}, 146 (2006).

\bibitem{gpds} M. Burkardt, Phys. Rev. D {\bf 72}, 094020 (2005); B. Pasquini, M. Pincetti and S.
Boffi, Phys. Rev. D {\bf 72} 094029 (2005).

\bibitem{soffer02} J. Soffer, M. Stratmann and W. Vogelsang, Phys. Rev. D {\bf 65}, 114024
(2002).

\bibitem{collins06} J.C. Collins {\it et al.}, Phys. Rev. D {\bf 73}, 094023
(2006).

\bibitem{anselmino03} M. Anselmino, U. D'Alesio and F. Murgia, Phys. Rev. D {\bf 67},
074010 (2003).

\end{thebibliography}
\end{document}